\documentstyle[12pt]{article}

\begin{document}
\title{
Pair creation of black holes \\ joined by cosmic strings}
\author{ R. Emparan \thanks{wmbemgar@lg.ehu.es} \\ Departamento de
F{\'\i}sica de la Materia Condensada\\ Universidad del Pa{\'\i}s Vasco,
Apdo. 644\\ 48080 Bilbao, Spain}
\date{}
\maketitle %
\setcounter{page}{0}
\pagestyle{empty}
\thispagestyle{empty}
\vskip 1.0in
\begin{abstract}
We argue that production of charged black hole
pairs joined by a cosmic string in the presence of a magnetic field can
be analyzed using the Ernst metric. The effect of the cosmic string is
to pull the black holes towards each other, opposing to the background
field. An estimation of the production rate using the Euclidean action
shows that the process is suppressed as compared to the formation of
black holes without strings.
\end{abstract}
\vfill

\begin{flushleft}
Phys.~Rev.~Lett.~{\bf 75}, 3386 (1995)\\EHU-FT 95/10\\
gr-qc/9506025\\ June 1995
\end{flushleft}

\newpage\pagestyle{plain} %

The idea that some gravitational instantons can be interpreted as
mediating tunnelling processes leading to, e.g., spontaneous formation
of black holes is an attractive one, and has provided some hints on
peculiarities of the yet-to-be-built quantum theory of gravity, such as
topology changing processes and the statistical properties of black
holes. The simplest example is provided by the Euclidean section of the
Schwarzschild metric, which was interpreted in Ref.\,\cite{gpy} as
yielding the nucleation rate of black holes in a thermal bath. More
recently there has been considerable interest in solutions that
describe the spontaneous creation of black hole pairs. The instantons
relevant here are obtained from the analytic continuation to Euclidean
time of geometries related to the $C$-metric, whose Lorentzian section
is known to represent a pair of black holes accelerating uniformly in
opposite directions \cite{kin}. In general, these solutions possess
conical singularities running along the axis in the direction joining
the black holes, either between the black holes or running from each
black hole to infinity. Physically, these singularities reflect the
presence of forces acting on the black holes. As a general rule, a
conical deficit tends to pull the black hole towards it, whereas a
conical excess pushes in the opposite direction.

An interesting modification of the $C$-metric is the Ernst metric
\cite{ernst}, obtained by adding a background electromagnetic field.
This can be adjusted so as to compensate exactly for the necessary
force to accelerate the black holes, and the resulting geometry is free
from conical singularities. This solution has been extensively studied
in recent years \cite{gib,gs,ggs,dggh,hhr}. However, as we will argue
below, this is not the only channel for the decay of the background
magnetic field. In fact, one can consider processes where the black
holes are joined by a cosmic string. It is then important to ascertain
whether the presence of the string enhances or suppresses the
probability for pair creation.

Below we will see that if, loosely speaking, the force exerted on the
black hole by the background field is in excess to the product of its
mass and acceleration then the Ernst metric has a conical deficit
running inbetween the black holes. We will interpret this physically by
regarding the conical deficit as created by a cosmic string which
decelerates the black holes\footnote{Very recently, another process
involving similar ingredients but a different physical phenomenon (the
breaking of a cosmic string to yield a pair of accelerating black
holes) has been considered in Ref.\,\cite{break}. It requires that the
conical deficit runs in directions opposed to what we consider here.}.
Clearly, a question to consider is whether it is consistent to
approximate a physical cosmic string  by a conical singularity even in
the vicinity of a black hole. This issue has been recently addressed in
Ref.\,\cite{anaetal}, where it is shown that a Nielsen-Olesen vortex
can effectively pierce a black hole (and, in fact, it is only slightly
distorted near the horizon) resulting in a geometry of a conical
singularity centered on a black hole. Moreover, in the same paper it is
argued that it is possible for a string to terminate on a black hole.
This can not be achieved in a topologically trivial space, since no
gauge can be taken that remains regular as we shrink to zero radius a
two-sphere connected to the string end. This topological obstruction
disappears when the vortex terminates in a black hole because the
spheres can not be contracted beyond the Schwarzschild radius. This
opens up the exciting possibility of considering a new variety of
string vortex-black hole interactions.

The black holes will be created in a background magnetic field,
described by the Melvin solution: \begin{eqnarray}\label{melvin}
ds^2&=&(1+{1\over 4} \widehat B_M^2 \rho^2)^2 (-dt^2 +dz^2+d\rho^2)
+{\rho^2 d\varphi^2\over (1 +{1\over 4} \widehat B_M^2 \rho^2)^2
}\nonumber\\ A_\varphi& =& {\rho^2 \widehat B_M\over 2(1+{1\over 4}
\widehat B_M^2 \rho^2)}  \end{eqnarray} The motion of a pair of charged
black holes uniformly accelerating in opposite directions in the
background of a magnetic field is represented in Einstein-Maxwell
theory by the Ernst solution:
\begin{eqnarray}\label{ernst} ds^2&=&
{\Lambda^2\over A^2 (x-y)^2} [G(y) dt^2 -G^{-1}(y) dy^2 + G^{-1}(x) dx^2
] \nonumber\\ &+& {G(x)\over \Lambda^2 A^2 (x-y)^2} d\varphi^2\,,\\
A_\varphi& =& -{2\over B \Lambda}\bigl(1+{1\over
2}Bqx\bigr)+k\,,\nonumber
\end{eqnarray}
where
\begin{eqnarray}
\Lambda&=& \bigl(1+{1\over 2}Bqx\bigr)^2+ {B^2\over 4
A^2(x-y)^2}G(x)\,,\nonumber\\
G(\xi)&=&(1+r_{-}A\xi)(1-\xi^2-r_{+}A\xi^3)\,,
\end{eqnarray} and
$q^2=r_- r_+$. For a detailed description of this geometry see, e.g.,
Ref.\,\cite{dggh}. Here we shall only summarize the most important
features.

The parameters in Eq.\,(\ref{ernst}) will be constrained so that
$G(\xi)$ has four real roots $\xi_1<\xi_2\leq\xi_3<\xi_4$, and we will
also set $\xi_1=-1/(r_-A)$. Then, the correct signature is obtained
when $x$ and $y$ are restricted to the ranges $\xi_3\leq x\leq\xi_4$,
$-\infty<y\leq x$. The surfaces $y=\xi_1, \xi_2, \xi_3$ correspond to
the inner black horizon, event black hole horizon and acceleration
horizon, respectively; $x=\xi_3, \xi_4$ are axes pointing towards
spatial infinity and to the other black hole, respectively. It is not
difficult to see that the parameter $A$ can loosely be thought of as
the acceleration of the black holes. We will also define $m = (r_+
+r_-)/2$, which can be identified as the mass of the black hole.
Another important feature is that the Ernst metric asymptotes to the
Melvin metric at spatial infinity ($x,y\rightarrow \xi_3$). Finally,
$k$ will be taken so as to confine Dirac string singularities to the
axis $x=\xi_4$.

It must be noted that $q$ and $B$ are not the physical magnetic charge
and field, but rather they approximate them in the limit $r_+
A, r_-A<<1$. In fact, the value of the magnetic field on the axis,
where it takes its maximum value, is $\widehat B=BG'(\xi_3)/2 L^{3/2}$
(we have defined $L\equiv\Lambda(\xi_3)$). Also, the adequate
definition of the physical charge of the black hole is $\widehat q
=1/4\pi \int F$, so that
\begin{equation}\label{qhat} \widehat q =q
{\Delta\varphi (\xi_4-\xi_3)\over 4\pi L^{1/2}(1+{1\over2}q B\xi_4)}\,,
\end{equation}
where $\Delta \varphi$ is the period of the azimuthal
coordinate $\varphi$.

The semiclassical approximation is expected to be reliable for small
values of $\widehat q \widehat B$. This will be equivalent to
considering small $r_+ A$. In this limit we find the following
expressions for the roots $\xi_i$:
\begin{eqnarray}\label{wfield}
\xi_2&=&-{1\over r_+ A}+r_+ A+\dots\,,\nonumber\\ \xi_3&=&-1-{r_+
A\over 2}+\dots\,,\\  \xi_4&=&1-{r_+ A\over 2}+\dots\nonumber
\end{eqnarray}

Now, in general, the Ernst metric contains conical singularities
running along the axes $x=\xi_3,\xi_4$. These can be avoided by
properly adjusting the period of $\varphi$. The situation we are
interested in requires absence of conical singularities at the axis
running to infinity. This is obtained by setting
\begin{equation}\label{fiper}
\Delta\varphi= {4\pi L^2\over
G'(\xi_3)}\,.
\end{equation}
If we also adjusted the value of the ratio
circumference/radius at $x=\xi_4$ to match the value (\ref{fiper}), we
would obtain the regular metric considered in \cite{ernst,gs,ggs,hhr}.
However, we have seen that it is also permissible to have a cosmic
string joining the black holes and creating a conical deficit inbetween
them. This requires
\begin{equation}\label{ineq}
G'(\xi_3)\Lambda^2(\xi_4)> -G'(\xi_4)L^2\,.
\end{equation}
In the
limiting case (\ref{wfield}), this inequality becomes approximately
$qB>m A$, indicating that the effect of the string is to oppose to the
force separating the black holes. The mass per unit length, $\mu$, of a
cosmic string is obtained as $1/8\pi$ times the conical deficit that it
creates. Then, in our case,
\begin{equation}\label{mu}
\mu={1\over
4}\biggl( 1-\left|{G'(\xi_4)L^2\over G'(\xi_3)\Lambda^2(\xi_4)}\right|
\biggr)\simeq qB-mA\,.
\end{equation}
With the choice (\ref{fiper}) the
physical magnetic field and charge in the weak field limit become
\begin{eqnarray}\label{phys}
\widehat B &\simeq& B(1- {1\over 2} qB
+2\mu)\,,\nonumber\\ \widehat q &\simeq& q(1-2\mu)\,.
\end{eqnarray}

The procedure to obtain an instanton mediating the decay process of the
magnetic field to a pair of black holes joined by a string has been
described in Ref.\,\cite{gs}. First, we continue $t=i\tau$ in the
metric (\ref{ernst}). Positive definiteness of the metric then requires
$\xi_2\leq y\leq \xi_3$. For non-extremal black holes, with $\xi_1\neq
\xi_2$, regularity at the horizons requires adjusting the surface
gravities and the period of $\tau$ (i.e., the inverse temperature
$\beta$) so that
\begin{equation}\label{tper}
\beta={4\pi \over
G'(\xi_3)}=-{4\pi \over G'(\xi_2)}\,.
\end{equation}
The second
equality is achieved when $\xi_1-\xi_2-\xi_3+\xi_4=0$, which imposes
the following restriction on the parameters:
\begin{equation}\label{restr}
A={r_+-r_-\over 2q^2}\sqrt{m\over r_-}\,.
\end{equation}
In the limit of small $r_+ A$, this means that the black
holes are close to extremality. Eqs.\,(\ref{mu},\ref{phys},\ref{restr})
can be used to express all the parameters in terms of $\widehat q$,
$\widehat B$ and $\mu$:
\begin{eqnarray}\label{interms}
r_\pm &\simeq &
\widehat q [1+2\mu \pm (\widehat q \widehat B -\mu)]\,,\nonumber\\  m
&\simeq& q \simeq \widehat q (1+2\mu) \,,\\ A &\simeq& \widehat B
[1+{1\over 2} \widehat q \widehat B-{\mu\over \widehat q \widehat B}
(1+2\widehat q \widehat B)]\,.\nonumber
\end{eqnarray}
Notice that the
effect of $\mu$ is to decrease the value of $A$, so the string
effectively decelerates the motion of the black holes. It is also
amusing to find that if we define, as in Ref.\,\cite{anaetal}, the
black hole internal energy  (or inertial mass) by $m_I\simeq m(1-2\mu)$
(we are taking into account that only half of the string is pinched on
the black hole), then $m_I=\widehat q$.

For extremal black holes, the event horizon is at infinite proper
distance and we need only match $\beta$ at the acceleration horizon
$y=\xi_3$.

Now we can slice the Euclidean solution in half along a constant $\tau$
surface (we have to take two antipodal values of $\tau$). The resulting
geometry is precisely that of the moment of closest approach of the
black holes in the Lorentzian Ernst metric. Moreover, since the
extrinsic curvature vanishes for both surfaces they can be glued
together. The process described in this way is the quantum tunnelling
from the Melvin metric to a pair of static black holes joined by a
string, that subsequently accelerate to infinity. If semiclassical
reasoning is valid, then the rate of the process in first approximation
should be given by $\exp(-I_{\rm cl})$, with $I_{\rm cl}$ the action of
the whole Euclidean solution. One could worry about the issue of the
finiteness of the quantum corrections to the decay rate. This is very
hard to analyze in a metric of a not very high degree of symmetry, like
Eq.\,(\ref{ernst}). Difficulties could be expected due to the infinite
redshift in the presence of horizons. However, the qualitative aspects
of this problem can be thoroughly analyzed in the much simpler case of
thermal nucleation of black holes using the Schwarzschild instanton
\cite{gpy}, and in this case it has been shown \cite{be} that
infinities can be properly renormalized in a low energy expansion of
the Einstein-Hilbert action.

The calculation of the Euclidean action can be performed along the same
lines as that for a pair of black holes without a string. In fact, one
can see that the derivation given in Refs.\,\cite{hhr} (see also
Ref.\,\cite{ggs,hh}) carries over directly, so we will only quote the
final result:
\begin{equation}\label{action}
I_{\rm cl}={2 \pi L^2
\over G'(\xi_3) A^2(\xi_3 -\xi_1)}\,.
\end{equation}
This result is
valid for both the extremal and non-extremal cases.

Now we must express the action in terms of the physical parameters,
which we do in the weak field limit:
\begin{equation}\label{finact}
I_{\rm cl} = \pi \widehat q^2 \biggl[ {1\over \widehat q \widehat B} -
{1\over2} + {\mu\over \widehat q \widehat B}\biggl(  {1\over \widehat q
\widehat B} +2 \biggr) +\dots\biggr] \,,
\end{equation}
for the
non-extremal case. The result for the extremal case can be obtained by
subtracting the contribution from the black hole area $-{\cal
A}_{\rm bh} /4 \simeq -\pi {\widehat q}^2$ from Eq.\,(\ref{finact})
\cite{hhr}.

We see in Eq.\,(\ref{finact}) that the action is increased when there
is a non-null string energy density $\mu$. Therefore, creation of black
holes joined by a cosmic string is suppressed relative to the case
where no string is present. In fact, we can give a simple heuristic
derivation of the contribution of the string to the action. If we
consider a particle of mass $\delta m$ at temperature $T$, its action
is given by $\delta I=\delta m/T$. Regard now the string as a
collection of particles of mass $\mu\times({\rm proper\, length})$,
i.e., $\delta m =\mu {\sqrt{|g_{yy}|}}_{x=\xi_4}\delta y$ at a local
inverse temperature $T^{-1}_y=\beta {\sqrt{|g_{tt}|}}_{x=\xi_4}$. Then
we calculate the string the action as
\begin{eqnarray}\label{istr}
I_{\rm string} &=& \mu \int_{\xi_2}^{\xi_3} dy
\beta\sqrt{|g_{tt}g_{yy}|}_{x= \xi_4}\nonumber\\ &\simeq& {\pi \mu\over
\widehat B^2}
\end{eqnarray}
which correctly reproduces the leading
contribution of the cosmic string to Eq.\,(\ref{finact}). Of course,
$I_{\rm string}$ is nothing but the Nambu-Goto action of the string
(the fact that vortices in the Abelian Higgs model are effectively
described by the Nambu-Goto action was arrived at independently in
Refs.\,\cite{fg}). Therefore we arrive at the simple result  that the
black hole pair production rate is modified in the semiclassical
approximation just by the nucleation rate of the cosmic string.

We have been considering that the black holes are pulled apart by an
electromagnetic field. Another force capable of accelerating black
holes is inflation. This requires considering the cosmological
$C$-metric, which has been studied in Ref.\,\cite{mr}. The results
qualitatively resemble those for the Ernst metric, so we can be quite
certain that the results presented here carry over to the cosmological
case without much difficulty.

\medskip

\section*{Acknowledgements} We are indebted to Ana Ach\'ucarro and Juan
Luis Ma{\~n}es for useful conversations and encouragement. This work
has been partially supported by a FPI grant from MEC (Spain) and
projects UPV 063.310-EB119-92 and CICYT AEN93-0435.

\end{document}